\documentclass[10pt,twocolumn,letterpaper]{article}

\usepackage{cvpr}              

\usepackage{graphicx}
\usepackage{amsmath}
\usepackage{amssymb}
\usepackage{booktabs}
\usepackage{pifont}
\usepackage{multirow}
\usepackage{makecell}

\usepackage{booktabs} 
\usepackage{array}    

\usepackage[pagebackref,breaklinks,colorlinks]{hyperref}

\usepackage[capitalize]{cleveref}
\crefname{section}{Sec.}{Secs.}
\Crefname{section}{Section}{Sections}
\Crefname{table}{Table}{Tables}
\crefname{table}{Tab.}{Tabs.}

\def\cvprPaperID{*****} 
\def\confName{CVPR}
\def\confYear{2022}

\begin{document}

\title {A-Eval: A Benchmark for Cross-Dataset Evaluation of Abdominal Multi-Organ Segmentation}

\author{
Ziyan Huang\textsuperscript{1,2}\thanks{Equal contribution. This work is done when Ziyan Huang is an intern at Shanghai AI Laboratory.}
\qquad Zhongying Deng\textsuperscript{2}\footnotemark[1]
\qquad Jin Ye\textsuperscript{2}\footnotemark[1]
\qquad Haoyu Wang\textsuperscript{2}\footnotemark[1]
\qquad Yanzhou Su\textsuperscript{2}
\\
Tianbin Li\textsuperscript{2}
\qquad Hui Sun\textsuperscript{2}
\qquad Junlong Cheng\textsuperscript{2}
\qquad Jianpin Chen\textsuperscript{2}
\\
Junjun He\textsuperscript{2}
\qquad Yun Gu\textsuperscript{1} 
\qquad Shaoting Zhang\textsuperscript{2} 
\qquad Lixu Gu\textsuperscript{1}\footnotemark[2]
\qquad Yu Qiao\textsuperscript{2}\thanks{Corresponding author}\\
\\
\textsuperscript{1}Shanghai Jiao Tong University\qquad \\ \textsuperscript{2}Shanghai AI Laboratory\\
{\tt\small \{ziyanhuang, gulixu\}@sjtu.edu.cn}\\
{\tt\small \{hejunjun, yejin, litianbin, zhangshaoting, qiaoyu\}@pjlab.org.cn}
}
\maketitle


\begin{abstract}
Although deep learning have revolutionized abdominal multi-organ segmentation,  models often struggle with generalization due to training on small, specific datasets. With the recent emergence of large-scale datasets, some important questions arise: \textbf{Can models trained on these datasets generalize well on different ones? If yes/no, how to further improve their generalizability?} 
To address these questions, we introduce A-Eval, a benchmark for the cross-dataset Evaluation ('Eval') of Abdominal ('A') multi-organ segmentation. We employ training sets from four large-scale public datasets: FLARE22, AMOS, WORD, and TotalSegmentator, each providing extensive labels for abdominal multi-organ segmentation. For evaluation, we incorporate the validation sets from these datasets along with the training set from the BTCV dataset, forming a robust benchmark comprising five distinct datasets. We evaluate the generalizability of various models using the A-Eval benchmark, with a focus on diverse data usage scenarios: training on individual datasets independently, utilizing unlabeled data via pseudo-labeling, mixing different modalities, and joint training across all available datasets. Additionally, we explore the impact of model sizes on cross-dataset generalizability. Through these analyses, we underline the importance of effective data usage in enhancing models' generalization capabilities, offering valuable insights for assembling large-scale datasets and improving training strategies. The code and pre-trained models are available at \href{https://github.com/uni-medical/A-Eval}{https://github.com/uni-medical/A-Eval}.
\end{abstract}

\begin{figure}[htbp]
\centering
\includegraphics[width=\columnwidth]{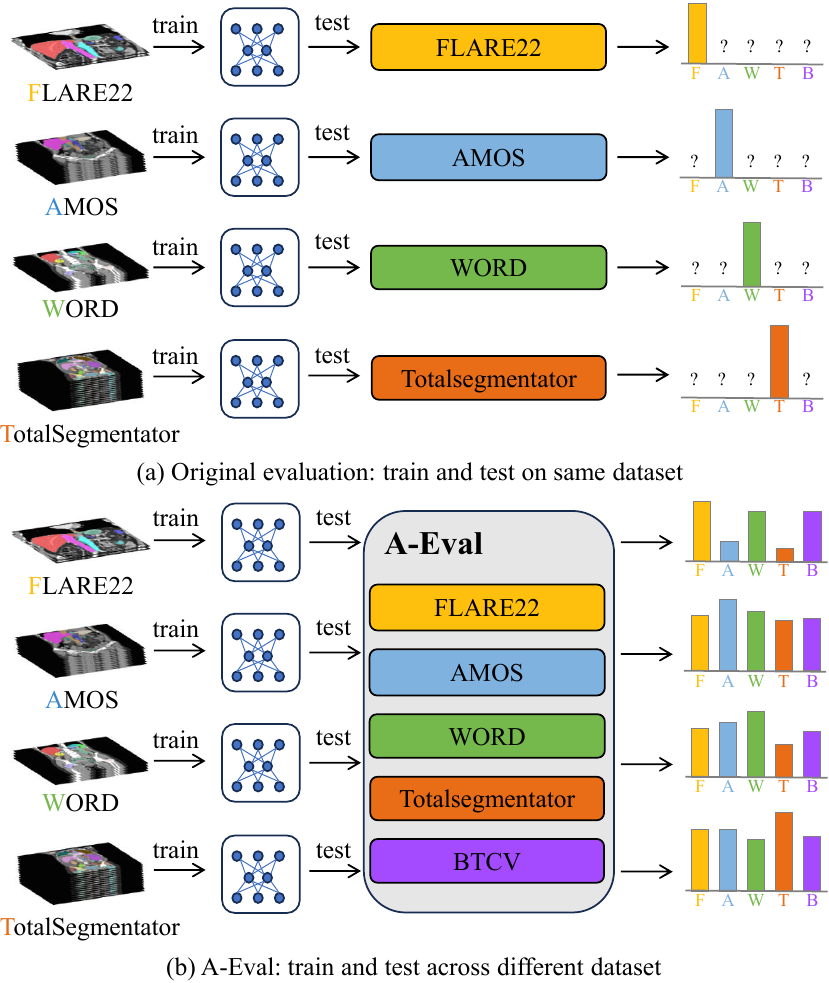}
\caption{Comparison of the original evaluation approach versus our proposed A-Eval benchmark for assessing model generalizability. (a) Original evaluation involves training and testing on the same dataset, providing good results but leaving uncertainty when applied to other datasets. (b) A-Eval, on the other hand, trains and tests across different datasets, offering a more comprehensive evaluation of model performance and its potential for generalizability.}
\label{fig:overview}
\end{figure}

\section{Introduction}
\label{sec:intro}

Accurate segmentation of abdominal organs is essential for clinical applications, especially in the diagnosis and treatment of prevalent abdominal cancers \cite{caner_statistics2022, tang2019clinically}. Traditionally, this labor-intensive and often tedious task has been manually carried out by specialists \cite{perez2022deep}. However, such a manual segmentation approach inevitably brings inaccurate results, particularly when imaging protocols and anatomical structures vary significantly in abdominal organs \cite{guo2020organ}. Deep learning has addressed this issue and revolutionized this field by introducing efficient and reliable methods \cite{gibson2018automatic, roth2018multi, wang2019abdominal, conze2021abdominal, pan2023abdomen}.

The success of deep learning in abdominal organ segmentation significantly relies on the quality and quantity of available training datasets \cite{van2019quality, tajbakhsh2020embracing}. 
Initially, the focus was mainly on the segmentation of individual organs along with their related tumors \cite{li2018hdenseunet, yu2019crossbar, man2019deep}. This trend was influenced by the constraints of early datasets such as MSD \cite{antonelli2022msd}, LiTS \cite{bilic2023lits}, and KiTS \cite{heller2019kits19, heller2023kits21}.
The advent of multi-organ datasets like BTCV \cite{landman2015btcv} allows for more holistic and complex abdominal studies, but their small size has limited their utility.
More recently, several large-scale datasets for abdominal organ segmentation, such as FLARE22 \cite{FLARE22}, AMOS \cite{ji2022amos}, WORD \cite{luo2022word}, and TotalSegmentator \cite{totalsegmentator} have emerged. These datasets are distinguished by their scale and the diversity of organs they include, thereby greatly expanding the possibilities for model development, refinement, and performance evaluation. However, while these models demonstrate impressive performance in segmenting abdominal organs within their original datasets, their generalizability across different datasets remains an open question, as shown in Figure \ref{fig:overview} (a).

The uncertainty in model generalizability can be attributed to several contributing factors, often referred to as 'domain gaps' or 'domain shifts' \cite{luo2022word, wang_embracing_2022, yan2019domain}. First, variations in imaging protocols across different medical centers introduce inconsistency. Second, a broad range of diseases represented in the training cohorts complicates the models' ability to generalize. Third, inconsistent image characteristics, such as contrast and resolution, create additional layers of variability. Finally, inconsistent annotation practices across different oncologists or radiologists further compromise the integrity of ground truth data. Although efforts have been made to include diverse data in these large abdominal datasets, their validation often remains confined to their own scope \cite{ji2022amos, totalsegmentator}. Even when some studies venture to test these models on external datasets, these efforts are often limited in scale and lack standardization, serving mainly as supplementary validations rather than comprehensive assessments \cite{FLARE22, luo2022word, ma2021abdomenct}. Such limitations restrict the full evaluation of model generalizability, highlighting the need for more comprehensive benchmarks to assess performance across diverse datasets.

To bridge this gap, we introduce A-Eval, a comprehensive benchmark specifically designed for cross-dataset evaluation in abdominal multi-organ segmentation. In A-Eval, 'A' signifies 'Abdomen,' and 'Eval' denotes 'Evaluation.' As illustrated in Figure \ref{fig:overview} (b), A-Eval incorporates the official training sets from four major public datasets: FLARE22 \cite{FLARE22}, AMOS \cite{ji2022amos}, WORD \cite{luo2022word}, and TotalSegmentator \cite{totalsegmentator}. Importantly, these datasets are large-scale and comprehensively cover multiple abdominal organs, ensuring a well-rounded evaluation of abdominal organ segmentation. Since the labels for some official test sets are not publicly available, we use the official validation sets from these four datasets for evaluation purposes and augment them with the training set from the BTCV \cite{landman2015btcv} dataset to enhance diversity. By combining these diverse collections, we establish a robust benchmark that includes five unique datasets. This design enables A-Eval to explicitly evaluate model generalizability across different datasets, offering a more comprehensive assessment than single-dataset benchmarks.

Based on the A-Eval benchmark, we conduct a thorough investigation into the factors affecting the ability of deep learning models to generalize in abdominal multi-organ segmentation. Initially, we train models on each of the four major datasets within A-Eval and test their performance across all five datasets. This approach offers a straightforward evaluation of how well models trained on existing large-scale datasets generalize. Subsequently, we delve into additional data-related factors influencing generalizability, including the utilization of unlabeled data from FLARE22 \cite{FLARE22}, handling multi-modality with CT and MR images from AMOS \cite{ji2022amos}, and the impact of joint training across multiple datasets. We also consider the role of model size in cross-dataset generalizability.
These analyses clarify how data usage and model architecture can improve performance and offer key insights for future work, making these models more reliable for real-world applications.

The main contributions of our study are two-fold:
\begin{enumerate}
\item We introduce A-Eval, a comprehensive benchmark designed for cross-dataset generalizability in abdominal multi-organ segmentation. The benchmark integrates training sets from FLARE22 \cite{FLARE22}, AMOS \cite{ji2022amos}, WORD \cite{luo2022word}, and TotalSegmentator \cite{totalsegmentator}, and employs their validation sets supplemented by BTCV \cite{landman2015btcv} for a robust evaluation across five distinct datasets.
\item Using A-Eval, we investigate model generalizability across  diverse data usage scenarios, including individual dataset training, unlabeled data utilization, multi-modality handling, and joint dataset training. We also explore the role of model size, providing key insights for enhancing generalizability and reliability in real-world settings.
\end{enumerate}

\section{Related Work}
\subsection{Abdominal Multi-Organ Segmentation Benchmarks}
Early benchmarks in abdominal organ segmentation primarily focused on individual organs and associated tumors. This focus is exemplified by datasets from the Medical Segmentation Decathlon (MSD) \cite{antonelli2022msd}, including MSD Liver, MSD Lung, MSD Pancreas, MSD Hepatic Vessel, MSD Spleen, and MSD Colon, as well as datasets like LiTS \cite{bilic2023lits}, KiTS \cite{heller2019kits19, heller2023kits21} and Pancreas-CT \cite{roth2015deeporgan}. The BTCV \cite{landman2015btcv} dataset, an initial step towards multi-organ segmentation, was constrained by its limited size. The CHAOS \cite{kavur2021chaos} dataset, although providing multi-modality segmentation from both CT and MRI data, also suffers from limited volume. 

Recent progress in the field has focused on the creation of numerous large-scale datasets for abdominal multi-organ segmentation \cite{rister2020ct, ma2021abdomenct, ma2022fast, FLARE22, luo2022word, totalsegmentator, qu2023annotating8000}. These datasets are characterized by their substantial volume and variety, encompassing numerous instances and diverse organ types. AMOS \cite{ji2022amos} stands out with its multimodal data inclusion. FLARE22 \cite{FLARE22} distinguishes itself by providing a small number of labeled cases and a substantially larger pool of unlabeled cases in its training set. Lastly, the TotalSegmentator dataset \cite{totalsegmentator} further extends the scope by offering full-body organ segmentation. 

Despite the growing diversity and size of abdominal multi-organ segmentation datasets, most existing benchmarks focus on intra-dataset evaluation, leaving the models' cross-dataset generalizability unexplored. Our work addresses this gap by introducing a benchmark explicitly designed for cross-dataset generalizability assessments.

\subsection{Model Generalizability}
Generalizability, the ability of a machine learning model to perform effectively on unseen data, is critical for medical image analysis models due to their anticipated applications in diverse real-world clinical scenarios \cite{yang2022genralizability, maleki2022generalizability, chan2020medicalreview}.
To enhance model generalizability, researchers typically pursue two main avenues: data-centric methods \cite{zha2023data_centric_review, jarrahi2022principles, zhu2017cyclegan, sandfort2019dataaug, chen2021dualct, su2023slaug} and model tweaks \cite{dou2018unsupervised, zhu2019boundary, liu2020ms}.

Data augmentation is a commonly used data-centric strategy \cite{jarrahi2022principles}.
Generative Adversarial Networks, specifically CycleGAN \cite{zhu2017cyclegan}, have been employed to augment CT data, leading to a significant boost in performance on non-contrast CT images \cite{sandfort2019dataaug}. Dual-Energy CT images and novel image fusion techniques have also been used to surpass traditional single-energy CT methods in segmentation accuracy, particularly in abdominal organs, thus boosting generalizability across different CT protocols and scanners\cite{chen2021dualct}. SLAug \cite{su2023slaug} employs class-level distributions and gradient guidance for enhanced data augmentation, reducing generalizability risk in unseen domains. 
Another simple but effective method to improve generalizability is not only increasing the volume but also enhancing the diversity and multi-center nature of training data \cite{ma2021abdomenct, FLARE22, ji2022amos}. However, the generalizability of models trained on these datasets is often only evaluated within the same dataset, lacking a cross-dataset evaluation.

On the architectural side, unsupervised domain adaptation frameworks have been developed that include specialized modules like Domain Adaptation (DAM) and Domain Critic Modules (DCM), aiming to improve cross-modality biomedical image segmentation \cite{dou2018unsupervised}.
Boundary-weighted domain adaptive networks like BOWDA-Net have been proposed to increase models' sensitivity to object boundaries in prostate MR images \cite{zhu2019boundary}.
MS-Net \cite{liu2020ms} incorporates Domain-Specific Batch Normalization layers and aggregates data from multiple sites, offering a robust solution for prostate segmentation in heterogeneous MRI data.

Unlike prior work focused on data augmentation or model tweaks, our study, using the A-Eval benchmark, centers on the impact of data diversity and model size on generalizability across multiple large-scale abdominal datasets.

\section{A-Eval Benchmark}
We present A-Eval, a benchmark explicitly aimed at standardizing the evaluation of abdominal organ segmentation across diverse large-scale datasets. This section details our approach to training, testing, data pre-processing, model architectures, and evaluation metrics.

\subsection{Datasets for A-Eval}
A-Eval incorporates five representative datasets, each carefully selected for its extensive scale, comprehensive organ annotations, diverse sources and diseases, as well as different image characteristics. This provides a reliable foundation for our evaluation. Details of these selected datasets are as follows: FLARE22 \cite{FLARE22}, AMOS \cite{ji2022amos}, WORD \cite{luo2022word}, TotalSegmentator \cite{totalsegmentator}, and BTCV \cite{landman2015btcv}. For model training, we employ the official training sets from the first four datasets. The corresponding official validation sets from these, along with the training set from BTCV, are used for evaluation.  A summary of these datasets, including their modalities, case numbers, the number of organ categories, and regions, is provided in Table \ref{tab:datasets}.

Some datasets encompass organ classes that others do not, as illustrated in Table \ref{tab:organs}. Therefore, to ensure a meaningful and fair comparison across datasets, we evaluate the models' performance based on a set of eight organ classes shared by all five datasets. These organ classes are \textit{liver, right kidney, left kidney, spleen, pancreas, gallbladder, esophagus, and stomach.} This selection enables a direct and consistent comparison across all datasets.

\begin{table*}[htbp!]
\centering
\begin{tabular}{lccccc}
\hline
\textbf{Dataset} & \textbf{Modality} & \textbf{\# Train} & \textbf{\# Test} & \textbf{\# Organs} & \textbf{Region} \\
\hline
FLARE22 \cite{FLARE22} 
& CT & \makecell[cc]{50 labeled \\ 2000 unlabeled} & 50 & 13 & \makecell[cc]{North American \\ European}  \\ 
AMOS \cite{ji2022amos}
& CT \& MR & \makecell[cc]{200 CT\\ 40 MR} & \makecell[cc]{100 CT \\ 20 MR} & 15 & Asian  \\ 
WORD \cite{luo2022word}
& CT & 100 & 20 & 16 & Asian   \\ 
TotalSegmentator \cite{totalsegmentator} 
& CT & 1082 & 57 & 104 & European   \\ 
BTCV \cite{landman2015btcv} 
& CT & - & 30 & 13 & North American \\ 
A-Eval Totals 
& CT \& MR & \makecell[cc]{1432 labeled CT \\ 2000 unlabeled CT \\ 40 MR} & \makecell[cc]{257 CT \\ 20 MR} & 8 & \makecell[cc]{North American \\ European \\ Asian} \\
\hline
\end{tabular}
\caption{Overview of the datasets used in A-Eval. The official training sets of FLARE22 \cite{FLARE22}, AMOS \cite{ji2022amos}, WORD \cite{luo2022word}, and TotalSegmentator \cite{totalsegmentator} are used for model training. Their official validation sets, along with the training set from BTCV \cite{landman2015btcv}, are employed as the test sets for model evaluation. The '\# Train' and '\# Test' columns indicate the number of labeled CT cases for training and testing, unless stated otherwise. The last row provides a cumulative summary of the datasets used in A-Eval.}
\label{tab:datasets}
\end{table*}

\begin{table*}[htbp!]
\centering
\resizebox{\textwidth}{!}{
\begin{tabular}{lcccccc}
\hline
\textbf{Organ Class} & \textbf{FLARE22 \cite{FLARE22}} & \textbf{AMOS \cite{ji2022amos}} & \textbf{WORD \cite{luo2022word}} & \textbf{TotalSegmentator \cite{totalsegmentator}} & \textbf{BTCV \cite{landman2015btcv}} & \textbf{A-Eval} \\
\hline
Liver & $\surd$ & $\surd$ & $\surd$ & $\surd$ & $\surd$ & $\surd$ \\
Kidney Right & $\surd$ & $\surd$ & $\surd$ & $\surd$ & $\surd$ & $\surd$ \\
Kidney Left & $\surd$ & $\surd$ & $\surd$ & $\surd$ & $\surd$ & $\surd$ \\
Spleen & $\surd$ & $\surd$ & $\surd$ & $\surd$ & $\surd$ & $\surd$ \\
Pancreas & $\surd$ & $\surd$ & $\surd$ & $\surd$ & $\surd$ & $\surd$ \\
Aorta & $\surd$ & $\surd$ & $\times$ & $\surd$ & $\surd$ & $\times$ \\
Inferior Vena Cava & $\surd$ & $\surd$ & $\times$ & $\surd$ & $\surd$ & $\times$ \\
Adrenal Gland Right & $\surd$ & $\surd$ & $\times$ & $\surd$ & $\surd$ & $\times$ \\
Adrenal Gland Left & $\surd$ & $\surd$ & $\times$ & $\surd$ & $\surd$ & $\times$ \\
Gallbladder & $\surd$ & $\surd$ & $\surd$ & $\surd$ & $\surd$ & $\surd$ \\
Esophagus & $\surd$ & $\surd$ & $\surd$ & $\surd$ & $\surd$ & $\surd$ \\
Stomach & $\surd$ & $\surd$ & $\surd$ & $\surd$ & $\surd$ & $\surd$ \\
Duodenum & $\surd$ & $\surd$ & $\surd$ & $\surd$ & $\times$ & $\times$ \\
\hline
\end{tabular}
}
\caption{Comparison of the presence of 13 organ classes across five different datasets and their intersection in our A-Eval evaluation, based on the FLARE22 \cite{FLARE22} label system. Each dataset column indicates whether a particular organ class is present ("$\surd$") or absent ("$\times$"). Notably, the WORD \cite{luo2022word} dataset includes a general annotation for adrenal glands without distinguishing between left and right. The 'A-Eval' column provides a summary, highlighting the organ classes that are consistently present across all datasets and are thus included in our A-Eval evaluation.}

\label{tab:organs}
\end{table*}

\subsection{Cross-Dataset Protocols}

To comprehensively evaluate the generalizability of abdominal multi-organ segmentation models, we define a set of data usage scenarios termed as cross-dataset protocols.
These protocols are designed to reflect diverse data scenarios commonly encountered in real-world applications.
These protocols include:

\textbf{Protocol 1. Training on Individual Dataset:} Initially, we train separate models on each dataset, focusing on utilizing the labeled CT data available in each one. This serves as a baseline for assessing the generalizability of models trained on individual large-scale datasets.

\textbf{Protocol 2. Using Unlabeled Data:} FLARE22 \cite{FLARE22} provides a set of unlabeled scans in addition to its labeled data. Through the assignment of pseudo-labels to these scans, we aim to investigate the impact of utilizing unlabeled data on model generalizability.

\textbf{Protocol 3. Using Multi-modal Data:} AMOS \cite{ji2022amos} offers both CT and MR scans, providing a unique opportunity to explore multi-modal training. We evaluate the effects of training models using exclusively CT scans, exclusively MR scans, and a combination of both.

\textbf{Protocol 4. Joint Training Across Datasets:} As the most comprehensive approach, we train a unified model across all available datasets. This enables us to assess the model's ability to generalize across diverse, large-scale datasets.

By exploring these protocols within our A-Eval, we aim to examine cross-dataset generalizability in abdominal multi-organ segmentation and highlight the importance of diverse dataset usage.

\subsection{Model Architecture and Training Procedure}

To ensure a fair comparison during our exploration of different data usage strategies, we consistently use the STU-Net \cite{huang2023stu} model architecture, a scalable and transferable derivative of the nnU-Net \cite{isensee2021nnu} specifically designed for medical image segmentation tasks.

STU-Net retains the fundamental symmetric encoder-decoder structure of nnU-Net, each containing residual blocks \cite{he2016deep} as their basic units. Each residual block is composed of two Conv-IN-LeakyReLU layers. The model incorporates six resolution stages and isotropic kernels (3,3,3) for all tasks. These features enhance the model's scalability and transferability. To avoid weight mismatch during task transfers, the model employs a downsampling operation within the first residual block of each stage and uses weight-free interpolation for upsampling.

In most of our experiments, we specifically employ STU-Net-L. This model variant was chosen due to its substantial parameter capacity, approximately 440MB, which ensures that our evaluations are not limited by model size. To investigate how the model size impacts cross-dataset generalizability, we study four STU-Net variants. These variants cover a range of parameter scales, from a compact 14M to a substantial 1.4B.

For preprocessing, each image is standardized to the same spacing via resampling, the value of which is automatically determined based on the dataset. We use different normalization methods depending on the modality: CT images are first clipped to a predetermined intensity range and then normalized using the dataset's mean and standard deviation values, whereas MR images and mixed CT and MR images undergo intensity normalization based on the mean and standard deviation calculated for each image.

The training process adheres to the standard nnU-Net pipeline, which incorporates on-the-fly data augmentation techniques, such as random rotations, scaling, and mirroring. The loss function is a combination of Dice and cross-entropy losses \cite{ma2021loss}.  Furthermore, to ensure convergence on the corresponding datasets, we adjust the number of training epochs according to the different data usage protocols we employ. This adaptive training strategy further enhances the fairness and robustness of our evaluation and comparison process.

\subsection{Evaluation Metrics and Inference Procedure}

For the evaluation of model performance across different cross-dataset protocols, we rely on two robust evaluation metrics -- Dice Similarity Coefficient (DSC) and Normalized Surface Dice (NSD). These metrics offer complementary perspectives on model performance, with DSC quantifying the overlap between predicted and ground-truth segmentations, and NSD providing a measure of agreement between the predicted and ground-truth boundaries.

The DSC is defined as follows:

\begin{equation}
DSC = \frac{2|P \cap G|}{|P| + |G|}
\end{equation}

Where \(P\) is the predicted segmentation, and \(G\) is the ground-truth segmentation. DSC values range between 0 and 1, with higher values indicating better performance.

The NSD is defined as:

\begin{equation}
\text{NSD}(G, S) = \frac{|B_{\partial G}(\tau) \cap \partial S| + |B_{\partial S}(\tau) \cap \partial G|}{|\partial G| + |\partial S|}
\end{equation}

Where \( B_{\partial G}(\tau) \) and \( B_{\partial S}(\tau) \) denote the border regions of the ground truth \( G \) and the segmentation surface \( S \) at a tolerance \( \tau \), respectively. \( \partial G \) and \( \partial S \) represent the boundaries of the ground truth and the segmentation. \( \tau \) is a tolerance defined based on clinical requirements or consistency between radiologists \cite{FLARE22}.

During inference, we follow the standard nnU-Net framework, utilizing a sliding window approach with a step size of 0.5. This approach ensures comprehensive coverage of the input volume, thereby allowing the model to exploit all available information. To further enhance evaluation robustness, we employ test-time augmentation (TTA) involving mirroring across the sagittal, coronal, and axial planes.

\begin{figure*}[htbp]
\centering
\includegraphics[width=\linewidth]{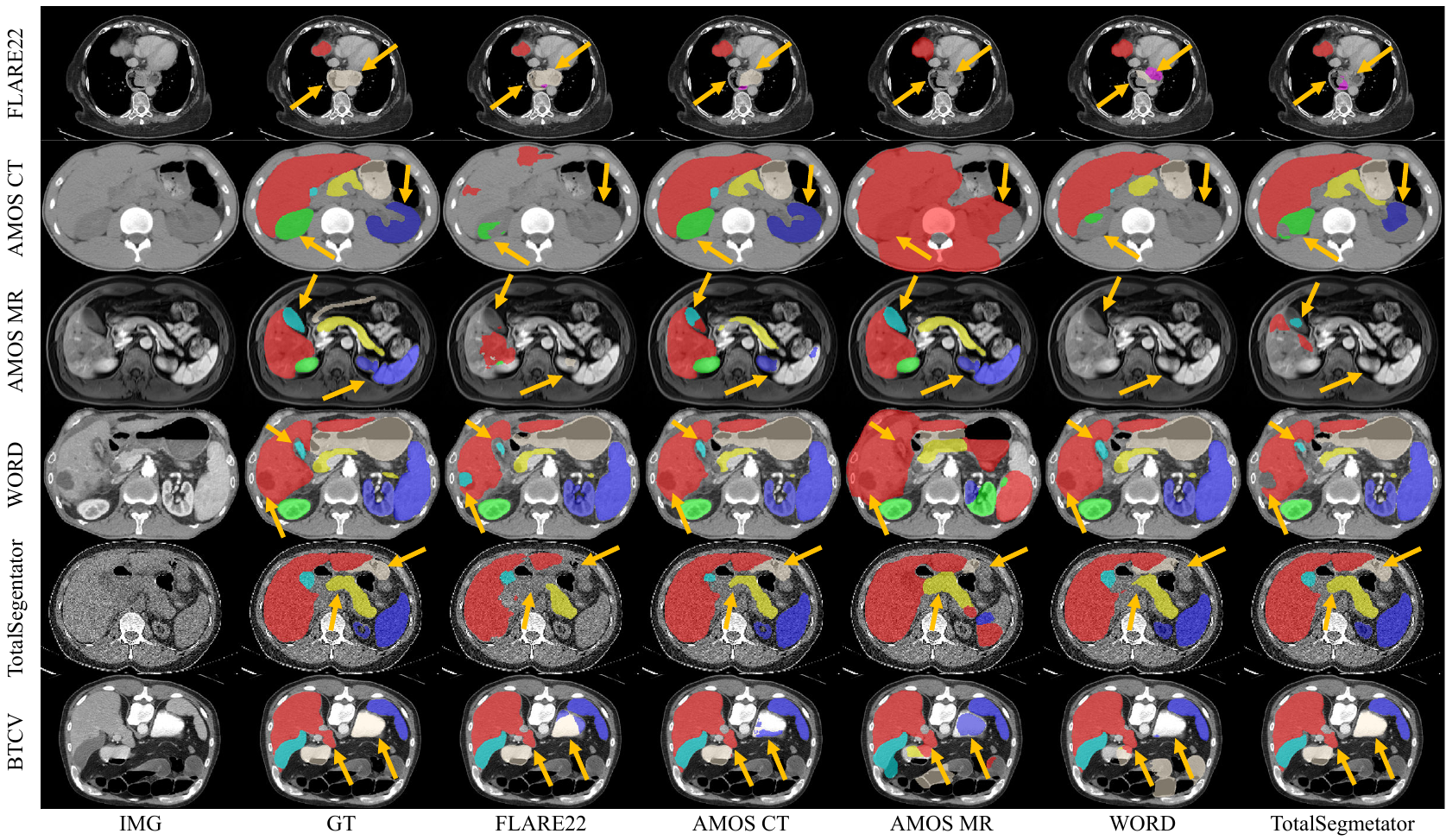}
\caption{Visualization of the performance of STU-Net-L models trained individually on different datasets (FLARE22 \cite{FLARE22}, AMOS CT \cite{ji2022amos}, AMOS MR \cite{ji2022amos}, WORD \cite{luo2022word}, TotalSegmentator \cite{totalsegmentator}) and validated on multiple datasets (FLARE22 \cite{FLARE22}, AMOS CT \cite{ji2022amos}, AMOS MR \cite{ji2022amos}, WORD \cite{luo2022word}, TotalSegmentator \cite{totalsegmentator}, BTCV \cite{landman2015btcv}) within the A-Eval Benchmark. Each row corresponds to testing on a different dataset, while each column depicts various elements: the original image, ground truth, and the segmentation results obtained from models trained individually on different datasets.}
\label{fig:visualization}
\end{figure*}

\begin{table*}[ht]
\centering
\resizebox{\textwidth}{!}{
\begin{tabular}{@{} l c c c c c c c c c c c c @{}}
\toprule
\multirow{2}{*}{\centering Train/Test} & \multicolumn{2}{c}{FLARE22} & \multicolumn{2}{c}{AMOS CT} & \multicolumn{2}{c}{WORD} & \multicolumn{2}{c}{TotalSegentator} & \multicolumn{2}{c}{BTCV} & \multicolumn{2}{c}{Mean $\pm$ SD} \\
\cmidrule(lr){2-3} \cmidrule(lr){4-5} \cmidrule(lr){6-7} \cmidrule(lr){8-9} \cmidrule(lr){10-11} \cmidrule(lr){12-13}
& DSC & NSD & DSC & NSD & DSC & NSD & DSC & NSD & DSC & NSD & DSC & NSD \\
\midrule
FLARE22
& 89.20 & 90.19 & 76.53 & 80.25 & 85.94 & 90.76 & 74.06 & 76.56 & 86.11 & 89.28 & 82.37 $\pm$ 5.94 & 85.41 $\pm$ 5.85 \\
AMOS CT
& 89.14 & 89.49 & \textbf{93.02} & \textbf{96.47} & 89.01 & 94.82 & 86.39 & 89.28 & 86.84 & 91.65 & 88.88 $\pm$ 2.35 & 92.34 $\pm$ 2.87 \\
WORD
& 86.86 & 88.73 & 87.53 & 92.34 & \textbf{90.92} & \textbf{95.75} & 80.58 & 83.47 & 84.69 & 88.74 & 86.12 $\pm$ 3.42 & 89.81 $\pm$ 4.10 \\
TotalSegentator
& \textbf{90.32} & \textbf{91.96} & 89.65 & 94.02 & 86.30 & 92.46 & \textbf{95.12} & \textbf{97.33} & \textbf{87.73} & \textbf{92.72} & \textbf{89.82 $\pm$ 3.00} & \textbf{93.70 $\pm$ 1.94} \\
\midrule
Mean $\pm$ SD
& 88.88 $\pm$ 1.26 & 90.09 $\pm$ 1.20 & 86.68 $\pm$ 6.18 & 90.77 $\pm$ 6.25 & 88.04 $\pm$ 2.04 & 93.45 $\pm$ 1.96 & 84.04 $\pm$ 7.74 & 86.66 $\pm$ 7.63 & 86.34 $\pm$ 1.11 & 90.60 $\pm$ 1.64 & 86.80 $\pm$ 4.88 & 90.31 $\pm$ 5.07 \\
\bottomrule
\end{tabular}
}
\caption{Performance comparison of the STU-Net-L model trained individually on different datasets and evaluated across various CT datasets. Displayed values represent average DSC (\%) and NSD (\%) for the eight shared classes across these datasets. The 'Mean $\pm$ SD' values, summarized both per row and per column, reflect the model's variability across various training and testing conditions. The overall mean and standard deviation values at the bottom right corner provide a summary of the model's generalization capabilities.}
\label{tab:model_performance}
\end{table*}

\begin{table*}[ht]
\centering
\resizebox{\textwidth}{!}{
\begin{tabular}{@{} l c c c c c c c c c c c c @{}}
\toprule
\multirow{2}{*}{\centering Train/Test} & \multicolumn{2}{c}{FLARE22} & \multicolumn{2}{c}{AMOS CT} & \multicolumn{2}{c}{WORD} & \multicolumn{2}{c}{TotalSegentator} & \multicolumn{2}{c}{BTCV} & \multicolumn{2}{c}{Mean $\pm$ SD} \\
\cmidrule(lr){2-3} \cmidrule(lr){4-5} \cmidrule(lr){6-7} \cmidrule(lr){8-9} \cmidrule(lr){10-11} \cmidrule(lr){12-13}
& DSC & NSD & DSC & NSD & DSC & NSD & DSC & NSD & DSC & NSD & DSC & NSD \\
\midrule
FLARE22 w/o PL
& 89.20 & 90.19 & 76.53 & 80.25 & 85.94 & 90.76 & 74.06 & 76.56 & 86.11 & 89.28 & 82.37 $\pm$ 5.94 & 85.41 $\pm$ 5.85\\
FLARE22 w/ PL
& \textbf{91.98} & \textbf{93.46} & \textbf{87.53} & \textbf{90.92} & \textbf{87.15} & \textbf{92.01} & \textbf{85.55} & \textbf{88.29} & \textbf{87.35} & \textbf{90.94} & \textbf{87.91 $\pm$ 2.15} & \textbf{91.12 $\pm$ 1.69}\\
\bottomrule
\end{tabular}
}
\caption{Performance comparison of the STU-Net-L model trained with and without Pseudo Labeling (PL) on the FLARE22 dataset. Displayed values represent average DSC (\%) and NSD (\%) for the eight shared classes across various CT datasets. The 'Mean $\pm$ SD' values summarize the average performance and variability across all datasets for each training method.}
\label{tab:model_performance_PL}
\end{table*}

\begin{table*}[ht]
\centering
\resizebox{\textwidth}{!}{
\begin{tabular}{@{} l c c c c c c c c c c c c c c @{}}
\toprule
\multirow{2}{*}{\centering Train/Test} & \multicolumn{2}{c}{FLARE22} & \multicolumn{2}{c}{AMOS CT} & \multicolumn{2}{c}{AMOS MR} & \multicolumn{2}{c}{WORD} & \multicolumn{2}{c}{TotalSegentator} & \multicolumn{2}{c}{BTCV} & \multicolumn{2}{c}{Mean \( \pm \) SD} \\
\cmidrule(lr){2-3} \cmidrule(lr){4-5} \cmidrule(lr){6-7} \cmidrule(lr){8-9} \cmidrule(lr){10-11} \cmidrule(lr){12-13} \cmidrule(lr){14-15}
& DSC & NSD & DSC & NSD & DSC & NSD & DSC & NSD & DSC & NSD & DSC & NSD & DSC & NSD \\
\midrule
AMOS CT
& 89.14 & 89.49 & 93.02 & 96.47 & 70.08 & 72.92 & 89.01 & 94.82 & 86.39 & 89.28 & 86.84 & 91.66 & 85.75 \( \pm \) 7.33 & 89.11 \( \pm \) 7.70 \\
AMOS MR
& 61.47 & 58.97 & 73.97 & 48.69 & 91.73 & 95.22 & 45.30 & 43.93 & 48.08 & 48.09 & 77.60 & 61.61 & 66.36 \( \pm \) 16.48 & 59.42 \( \pm \) 17.18 \\
AMOS CT+MR
& \textbf{89.81} & \textbf{90.46} & \textbf{93.24} & \textbf{96.80} & \textbf{92.72} & \textbf{96.58} & \textbf{89.36} & \textbf{95.18} & \textbf{88.42} & \textbf{91.36} & \textbf{87.66} & \textbf{92.53} & \textbf{90.20 \( \pm \) 2.08} & \textbf{93.82 \( \pm \) 2.50} \\
\bottomrule
\end{tabular}
}
\caption{Performance comparison of the STU-Net-L model trained on CT, MR, and combined CT+MR modalities of the AMOS dataset across various datasets. Displayed values represent average DSC (\%) and NSD (\%) for the eight shared classes across these datasets. The 'Mean \( \pm \) SD' values provide a summary of the model's performance and variability across all datasets for each training method.}
\label{tab:model_performance_MR}
\end{table*}

\section{Experiments and Results}
\subsection{Implementation Details}
All the experiments were conducted in a CentOS 7 environment using Python 3.9 and PyTorch 2.0 with nnU-Net 2.1. We followed the default data preprocessing and training procedures provided by nnU-Net. Importantly, optimal patch sizes and target spacing were automatically configured by nnU-Net based on the characteristics of each training dataset. We employed the SGD optimizer, setting Nesterov momentum at 0.99 with a weight decay of 1e-3. 
We kept the batch size constant at 2 and designed each epoch to comprise 250 iterations. The initial learning rate was set at 0.01 for training from scratch and was adjusted over time according to the poly learning rate policy, expressed as $(1-\text{{epoch}}/\text{{total epochs}})^{0.9}$. 
Regarding the training duration, we adopted different strategies tailored to each dataset, acknowledging that the required training time for models to converge can vary across different datasets. In line with the practice suggested by TotalSegmentator \cite{totalsegmentator} and STU-Net \cite{huang2023stu}, we extended the training on the TotalSegmentator dataset to 4000 epochs, exploiting its substantial scale and complexity to the fullest. However, for the remaining datasets - FLARE22 \cite{FLARE22}, AMOS \cite{ji2022amos}, and WORD \cite{luo2022word}, we adhered to nnU-Net's default configuration of 1000 epochs, as our testing revealed no substantial performance gains beyond this point.

\subsection{Cross-Dataset Evaluation for Models Trained on Individual Datasets}
In this section, we assess the generalizability of models that are individually trained on FLARE22 \cite{FLARE22}, AMOS CT \cite{ji2022amos}, WORD \cite{luo2022word}, and TotalSegmentator \cite{totalsegmentator} datasets (Protocol 1). It is important to note that only labeled CT data were used for FLARE22 \cite{FLARE22} and exclusively CT data for AMOS \cite{ji2022amos}. 

As shown in Table \ref{tab:model_performance}, there exists significant variation in model performance across different datasets. When examining the datasets from a training perspective, it becomes evident that models trained on individual datasets generally perform best when tested on the same datasets but often underperform when evaluated on others. These trends are visually depicted in Figure \ref{fig:visualization}.
Overall, the model trained on the TotalSegmentator \cite{totalsegmentator} dataset achieves the highest average performance (DSC of 89.82\% and NSD of 93.70\%) with the lowest standard deviations (SD for DSC is 3.00\% and for NSD is 1.94\%). This indicates superior generalizability and stability of this model across different test datasets. In contrast, the FLARE22-trained \cite{FLARE22} model lags behind, scoring lower in both average performance and stability (DSC of 82.37\%, NSD of 85.41\%, SD for DSC is 5.94\%, and for NSD is 5.85\%). 
Interestingly, there seems to be a direct correlation between the size of the training dataset and the model's generalizability, following the order of FLARE22 \textless WORD \textless AMOS CT \textless TotalSegmentator. This observation reveals that larger training datasets enhance the model's generalizability.

From a testing perspective, the BTCV dataset shows low variability (SD for DSC of 1.11\% and for NSD of 1.64\%), indicating consistent but less discriminating evaluations. In contrast, the TotalSegmentator and AMOS CT datasets, show greater performance variability: 7.74\% for DSC and 7.63\% for NSD in TotalSegmentator, and 6.18\% for DSC and 6.25\% for NSD in AMOS CT. This suggests these datasets contain unique, challenging validation samples that resist easy generalization. For example, some TotalSegmentator validation cases may exclude the abdominal region, raising the bar for model generalization.

\begin{table*}[ht]
\centering
\resizebox{\textwidth}{!}{
\begin{tabular}{@{} l c c c c c c c c c c c c @{}}
\toprule
\multirow{2}{*}{\centering Train/Test} & \multicolumn{2}{c}{FLARE22} & \multicolumn{2}{c}{AMOS CT} & \multicolumn{2}{c}{AMOS MR} & \multicolumn{2}{c}{WORD} & \multicolumn{2}{c}{TotalSegentator} & \multicolumn{2}{c}{Mean \( \pm \) SD} \\
\cmidrule(lr){2-3} \cmidrule(lr){4-5} \cmidrule(lr){6-7} \cmidrule(lr){8-9} \cmidrule(lr){10-11} \cmidrule(lr){12-13}
& DSC & NSD & DSC & NSD & DSC & NSD & DSC & NSD & DSC & NSD & DSC & NSD \\
\midrule
Individual Train
& 89.20 & 90.19 & 93.02 & 96.47 & 91.73 & 95.22 & 90.92 & 95.75 & 95.12 & 97.33 & 92.00 \( \pm \) 1.99 & 94.99 \( \pm \) 2.50 \\
Joint Train
& 91.98 & 93.58 & 92.42 & 96.42 & 90.87 & 95.28 & 88.88 & 94.20 & 93.87 & 96.10 & 91.60 \( \pm \) 1.67 & 95.12 \( \pm \) 1.09 \\
\bottomrule
\end{tabular}
}
\caption{Comparison of the STU-Net-L model's performance on various datasets when trained individually on the corresponding dataset (Individual Train) versus when jointly trained on all labeled datasets (Joint Train). Displayed values represent average DSC (\%) and NSD (\%) for the eight shared classes across these datasets. The 'Mean \( \pm \) SD' values show the model's performance and corresponding variability across different training methods and datasets.}
\label{tab:model_performance_Joint_Train}
\end{table*}

\subsection{Impact of Pseudo-Labeling on Model Generalizability}
This section centers around Protocol 2, which can be used to investigate the influence of the pseudo-labeling (PL) technique on the generalization capability of models. Protocol 2 is with a particular focus on the FLARE22 dataset, which comprises 2000 unlabeled and 50 labeled images.

Following the champion solution of FLARE22 \cite{huang2022revisiting}, we first train STU-Net-L model exclusively on the 50 labeled images from the FLARE22 dataset. This model was then used to generate pseudo-labels for the unlabeled images, thereby creating an augmented dataset inclusive of these new labels.  In this process, we do not conduct label selection to maintain simplicity.
Subsequently, we retrained the model on this augmented dataset to examine whether the unlabeled data could enhance the model's generalization ability.

As presented in Table \ref{tab:model_performance_PL}, the utilization of pseudo-labeling (PL) demonstrates a clear improvement in the model's generalization capability. The mean DSC increased from 82.37\% (without PL) to 87.91\% (with PL), while the mean NSD increased from 85.41\% to 91.12\%. Concurrently, the standard deviations of DSC and NSD decrease from 5.94\% and 5.85\% (without PL) to 2.15\% and 1.69\% (with PL) respectively. These changes not only indicate a boost in performance but also an increase in stability across various datasets. The significant and consistent improvement across all datasets underscores the efficacy of pseudo labeling in enhancing model performance.

\subsection{Impact of Multi-Modality Data on Model Generalizability}
In this section, we delve into the Protocol 3 of multi-modality training to evaluate the cross-modality generalizability of existing models. 
For this purpose, we utilize the AMOS \cite{ji2022amos} dataset, comprising both CT and MR data. We explore three training scenarios: using only the 100 available CT images, only the 40 MR images, and a combination of both modalities. To ensure a comprehensive assessment of cross-modality generalizability, we integrate the official AMOS MR validation set into our cross-dataset evaluation.

As demonstrated in Table \ref{tab:model_performance_MR} and Figure \ref{fig:visualization}, models trained solely on one modality exhibit limitations when evaluated on the other. Specifically, a model trained exclusively on AMOS CT data excels in other CT datasets but falls short on the AMOS MR dataset, with a DSC of 70.08\% and an NSD of 72.92\%. Conversely, a model trained only on AMOS MR data performs well on its own MR validation set but poorly on CT datasets. Notably, a model trained on both CT and MR data  from the AMOS dataset (AMOS CT+MR) outperforms those trained on individual modalities across all datasets. This result highlights the utility of simple mixed-modality training in multi-modal learning for medical image segmentation and suggests the potential benefits of integrating MR data into traditionally CT-focused training procedures.

\subsection{Improving Generalizability Through Joint Training Across Multiple  Datasets}
In this section, we explore the improvement of model generalizability through joint training across multiple datasets, as per Protocol 4. We employ a jointly trained model that utilizes a comprehensive set of 3472 images available in the A-Eval benchmark. This set includes 1432 labeled CT images, 2000 unlabeled CT images, and 40 labeled MR images, as detailed in Table \ref{tab:datasets}. We compare the performance of this jointly trained model to models trained on individual datasets, evaluated on their respective validation sets. Additionally, we assess its ability to generalize using the BTCV dataset, which was excluded from the training process.

Initially, we train a model on labeled data from three datasets: FLARE22, AMOS, and TotalSegmentator, with labels that are aligned to match the FLARE22 label system. Using this pre-trained model, we generate pseudo labels for the WORD dataset's four missing categories. A second round of training incorporates these newly annotated data. Finally, we create pseudo labels for 2000 unlabeled images with the updated model and use them in a last round of joint training from all datasets.

As shown in Table \ref{tab:model_performance_Joint_Train}, the model trained through joint training consistently matches or surpasses the performance of models trained on individual datasets. For example, it outperforms the FLARE22-specific model, achieving a DSC of 91.98\% vs. 89.20\% and NSD of 93.58\% vs. 90.19\%.  Notably, the joint-trained model also performs well on the AMOS MR dataset, even though MR images make up only 1.15\% of the total training set. This suggests that despite an imbalanced mix of imaging modalities, the model retains its effectiveness in processing MR data through joint training.

The efficacy of joint training is further validated on the untested BTCV dataset, as shown in Tables \ref{tab:model_performance_BTCV_DSC} and \ref{tab:model_performance_BTCV_NSD}. Despite BTCV being excluded from the training process, the joint model surpasses all models trained on individual datasets when tested on it, achieving the highest mean DSC (88.90\%) and NSD (93.81\%) scores. This indicates that joint training with multiple datasets improves the model's generalizability to unseen data, surpassing single-dataset models.

\begin{table*}[htbp]
\centering
\resizebox{\textwidth}{!}{
\begin{tabular}{@{} l *{8}{c} c @{}}
\toprule
Train data & Liver & Right Kidney & Spleen & Pancreas & Gallbladder & Esophagus & Stomach & Left Kidney &  Mean \\
\midrule
FLARE22
& 94.48 & 86.68 & 91.27 & 82.00 & 76.31 & 81.29 & 89.31 & 87.51 & 86.11 \\
FLARE22 w/ PL
& 95.33 & 88.29 & 92.00 & \textbf{83.90} & 78.17 & \textbf{82.40} & 90.88 & 87.81 & 87.35 \\
AMOS CT
& 95.47 & 89.74 & 93.24 & 82.64 & 76.92 & 81.35 & 87.03 & 88.37 & 86.84 \\
AMOS CT+MR
& 95.59 & 90.10 & 92.28 & 82.04 & \textbf{79.97} & 81.72 & 90.57 & 89.00 & 87.66 \\
WORD
& 94.24 & 88.04 & 91.61 & 80.26 & 73.71 & 79.51 & 83.25 & 86.93 & 84.69 \\
TotalSegmentator
& 96.29 & 91.38 & 93.24 & 81.98 & 73.65 & 79.73 & 92.03 & \textbf{93.54} & 87.73 \\
\midrule
Joint Train
& \textbf{96.51} & \textbf{92.06} & \textbf{94.48} & 83.40 & 78.36 & 81.11 &\textbf{93.15} & 92.11 & \textbf{88.90} \\
\bottomrule
\end{tabular}
}
\caption{The DSC (\%) scores of the STU-Net-L models trained individually or jointly on all datasets for the shared eight organ categories in the BTCV dataset. The 'Mean' column denotes the average DSC across all organ categories.}
\label{tab:model_performance_BTCV_DSC}
\end{table*}

\begin{table*}[htbp]
\centering
\resizebox{\textwidth}{!}{
\begin{tabular}{@{} l *{8}{c} c @{}}
\toprule
Train data & Liver & Right Kidney & Spleen & Pancreas & Gallbladder & Esophagus & Stomach & Left Kidney &  Mean \\
\midrule
FLARE22
& 94.43 & 85.39 & 93.40 & 94.19 & 76.80 & 91.03 & 91.98 & 87.02 & 89.28 \\
FLARE22 w/ PL
& 95.57 & 87.30 & 94.54 & \textbf{95.75} & 79.84 & 92.37 & 95.25 & 86.89 & 90.94 \\
AMOS CT
& 96.55 & 93.09 & 95.35 & 93.93 & 78.63 & 92.42 & 91.67 & 91.54 & 91.65 \\
AMOS CT+MR
& 96.90 & 94.07 & 94.37 & 93.06 & \textbf{81.56} & \textbf{93.03} & 94.42 & 92.80 & 92.53 \\
WORD
& 93.26 & 90.09 & 93.18 & 93.39 & 73.34 & 91.35 & 87.35 & 87.95 & 88.74 \\
TotalSegentator
& 97.82 & \textbf{95.48} & 94.98 & 94.08 & 74.00 & 91.93 & 96.01 & \textbf{97.47} & 92.72 \\
\midrule
Joint Train
& \textbf{97.83} & 95.44 & \textbf{96.92} & 94.75 & 80.18 & 92.56 & \textbf{97.08} & 95.68 & \textbf{93.81} \\
\bottomrule
\end{tabular}
}
\caption{The NSD (\%) scores of the STU-Net-L models trained individually or jointly on all datasets for the shared eight organ categories in the BTCV dataset. The 'Mean' column denotes the average NSD across all organ categories.}
\label{tab:model_performance_BTCV_NSD}
\end{table*}

\begin{figure}[htbp]
\centering
\includegraphics[width=\columnwidth]{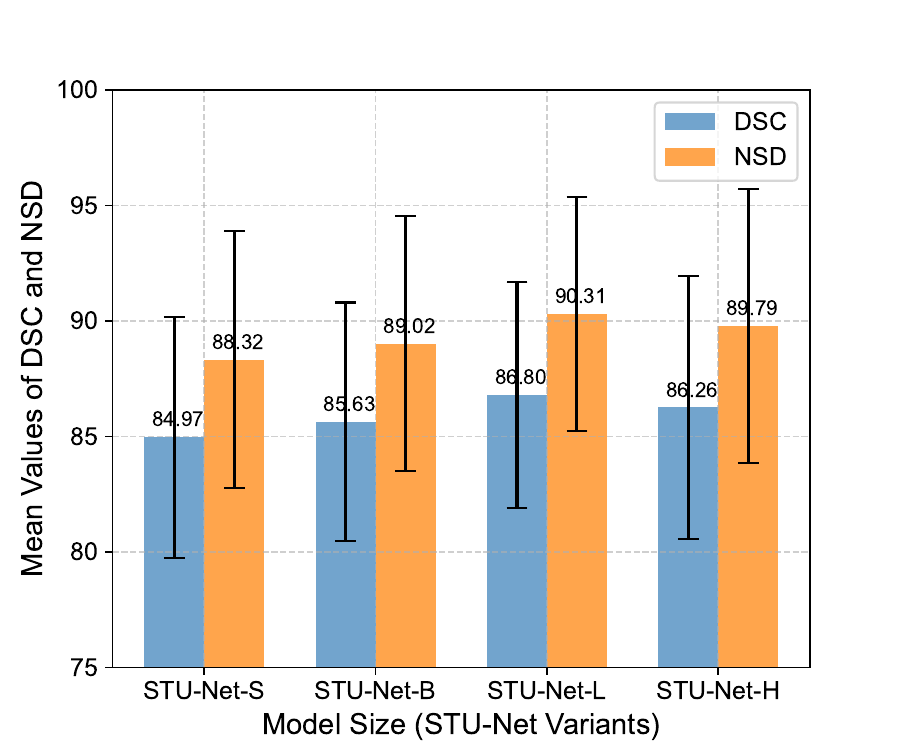}
\caption{Comparison of generalizability for STU-Net models of different sizes. Blue and red bars denote mean DSC and NSD values, respectively. Means are calculated from 20 cross-dataset evaluations (trained on four datasets and tested on five). Error bars represent the standard deviation, indicating the model's generalizability variability.}
\label{fig:model_size}
\end{figure}

\subsection{Impact of Model Size on Generalizability}
In this section, we investigate the influence of varying model sizes on model generalizability. We study four unique variants of the STU-Net model: STU-Net-S, STU-Net-B, STU-Net-L, and STU-Net-H, which correspond to 'small', 'base', 'large', and 'huge' sizes respectively. These models span a range of parameter scales, from a compact 14M to a significant 1.4B.

We conduct cross-dataset generalizability tests on each variant, similar to those described for STU-Net-L in Table \ref{tab:model_performance}. Each model is trained on four datasets and evaluated on the other five, with the Dice Similarity Coefficient (DSC) and Normalized Surface Dice (NSD) calculated across eight shared classes. The mean and standard deviation (SD) of these 20 values represent each model's generalizability. Detailed results are provided in the Appendix for reference.

Figure \ref{fig:model_size} presents the average DSC and NSD values derived from all test datasets for each STU-Net variant. A clear trend is noticeable: as the model size increases from 'small' to 'base' to 'large', there is a steady rise in performance. For DSC values, there is a successive increase from 84.65 for STU-Net-S, to 85.41\% for STU-Net-B, and further to 86.64\% for STU-Net-L. A similar progression is detected in NSD values, with a rise from 88.29\% for STU-Net-S, to 89.12\% for STU-Net-B, and peaking at 90.14\% for STU-Net-L.

However, when we scale up the model to the 'huge' variant (STU-Net-H), the performance does not proportionally increase. Despite its size, STU-Net-H's DSC (86.16\%) and NSD (89.95\%) do not surpass the 'large' variant. This suggests that merely enlarging the model does not always improve generalizability and may cause overfitting, particularly when training data is limited in size and diversity.

\section{Conclusion}

In this paper, we introduce A-Eval, a rigorous benchmark specifically designed for evaluating the cross-dataset generalizability of models in abdominal multi-organ segmentation. A-Eval acts as a standardized platform for in-depth investigation into various factors affecting model generalizability, including model architecture, training strategies, and training data. By offering such a comprehensive framework, A-Eval serves as an invaluable blueprint for future investigations in the development of models with superior generalizability.

Our A-Eval-based findings underscore the critical importance of data-centric strategies for achieving superior model generalizability. We have found that utilizing larger training datasets, integrating unlabeled data via pseudo-labeling, employing multi-modal learning, and conducting joint training across multiple datasets significantly enhances both model performance and generalizability. Furthermore, our results suggest that appropriately increasing a model's size contributes to better performance, thereby highlighting the potential of larger models in achieving enhanced generalizability. These insights offer actionable recommendations for assembling future large-scale datasets and serve as a roadmap for the design of models with robust generalizability, laying the foundation for advancements in abdominal multi-organ segmentation and beyond.

{\small
\bibliographystyle{ieee_fullname}
\bibliography{egbib}
}
\newpage
\appendix

\begin{table*}[ht]
\centering
\resizebox{\textwidth}{!}{
\begin{tabular}{@{} l c c c c c c c c c c c c c c c c @{}}
\toprule
\multirow{2}{*}{\centering Train/Test} & \multicolumn{2}{c}{FLARE22} & \multicolumn{2}{c}{AMOS CT} & \multicolumn{2}{c}{WORD} & \multicolumn{2}{c}{TotalSegentator} & \multicolumn{2}{c}{BTCV} & \multicolumn{2}{c}{CT Mean $\pm$ SD} & \multicolumn{2}{c}{AMOS MR} & \multicolumn{2}{c}{All Mean $\pm$ SD} \\
\cmidrule(lr){2-3} \cmidrule(lr){4-5} \cmidrule(lr){6-7} \cmidrule(lr){8-9} \cmidrule(lr){10-11} \cmidrule(lr){12-13} \cmidrule(lr){14-15} \cmidrule(lr){16-17}
& DSC & NSD & DSC & NSD & DSC & NSD & DSC & NSD & DSC & NSD & DSC & NSD & DSC & NSD & DSC & NSD \\
\midrule
FLARE22 w/o PL
& 89.20 & 90.19 & 76.53 & 80.25 & 85.94 & 90.76 & 74.06 & 76.56 & 86.11 & 89.28 & 82.37 $\pm$ 5.94 & 85.41 $\pm$ 5.85 & 24.77 & 23.96 & 72.77 $\pm$ 22.14 & 75.17 $\pm$ 23.52 \\
FLARE22 w/ PL
& \textbf{91.98} & 93.46 & 87.53 & 90.92 & 87.15 & 92.01 & 85.55 & 88.29 & 87.35 & 90.94 & 87.91 $\pm$ 2.15 & 91.12 $\pm$ 1.69 & 42.74 & 44.19 & 80.38 $\pm$ 16.95 & 83.30 $\pm$ 17.56 \\
AMOS CT
& 89.14 & 89.49 & 93.02 & 96.47 & 89.01 & 94.82 & 86.39 & 89.28 & 86.84 & 91.65 & 88.88 $\pm$ 2.35 & 92.34 $\pm$ 2.87 & 70.08 & 72.92 & 85.75 $\pm$ 7.33 & 89.11 $\pm$ 7.70 \\
AMOS MR
& 61.47 & 59.97 & 73.97 & 48.69 & 45.30 & 43.93 & 48.08 & 48.09 & 77.60 & 61.61 & 61.28 $\pm$ 13.09 & 52.26 $\pm$ 6.81 & 91.73 & 95.22 & 66.36 $\pm$ 16.48 & 59.42 $\pm$ 17.18 \\
AMOS CT+MR
& 89.81 & 90.46 & \textbf{93.24} & \textbf{96.80} & 89.36 & 95.18 & 88.42 & 91.36 & 87.66 & 92.53 & 89.70 $\pm$ 1.92 & 93.27 $\pm$ 2.37 & \textbf{92.72} & \textbf{96.58} & 90.20 $\pm$ 2.08 & 93.82 $\pm$ 2.50 \\
WORD
& 86.86 & 88.73 & 87.53 & 92.34 & \textbf{90.92} & \textbf{95.75} & 80.58 & 83.47 & 84.69 & 88.74 & 86.12 $\pm$ 3.42 & 89.81 $\pm$ 4.10 & 27.38 & 30.75 & 76.33 $\pm$ 22.11 & 79.96 $\pm$ 22.33 \\
TotalSegmentator
& 90.32 & 91.96 & 89.65 & 94.02 & 86.30 & 92.46 & \textbf{95.12} & \textbf{97.33} & 87.73 & 92.72 & 89.82 $\pm$ 3.00 & 93.70 $\pm$ 1.94 & 38.72 & 40.44 & 81.31 $\pm$ 19.24 & 84.82 $\pm$ 19.93 \\
Joint Train
& \textbf{91.98} & \textbf{93.58} & 92.42 & 96.46 & 88.88 & 95.28 & 93.87 & 96.10 & \textbf{88.90} & \textbf{93.80} & \textbf{91.21 $\pm$ 1.99} & \textbf{95.04 $\pm$ 1.16} & 90.87 & 95.28 & \textbf{91.15 $\pm$ 1.83} & \textbf{95.08 $\pm$ 1.07} \\
\bottomrule
\end{tabular}
}
\caption{Comprehensive performance comparison of STU-Net-L trained on various datasets as described in the main text. Each method's performance is reported in terms of DSC (\%) and NSD (\%) scores. Given that most methods were trained on CT datasets, the 'Mean \(\pm\) SD' column separately provides the average and standard deviation of both DSC and NSD across the five CT datasets and the overall six datasets including AMOS MR.}
\label{tab:comprehensive_model_performance}
\end{table*}

\begin{table*}[htbp]
\centering
\resizebox{\textwidth}{!}{
\begin{tabular}{@{} l c c c c c c c c c c c c @{}}
\toprule
\multirow{2}{*}{\centering Train/Test} & \multicolumn{2}{c}{FLARE22} & \multicolumn{2}{c}{AMOS CT} & \multicolumn{2}{c}{WORD} & \multicolumn{2}{c}{TotalSegentator} & \multicolumn{2}{c}{BTCV} & \multicolumn{2}{c}{Mean $\pm$ SD} \\
\cmidrule(lr){2-3} \cmidrule(lr){4-5} \cmidrule(lr){6-7} \cmidrule(lr){8-9} \cmidrule(lr){10-11} \cmidrule(lr){12-13}
& DSC & NSD & DSC & NSD & DSC & NSD & DSC & NSD & DSC & NSD & DSC & NSD \\
\midrule
FLARE22
& 86.97 & 87.59 & 73.18 & 76.46 & 85.21 & 90.38 & 71.75 & 74.05 & 84.24 & 87.47 & 80.27 $\pm$ 6.45 & 83.19 $\pm$ 6.61 \\
AMOS CT
& 87.63 & 87.43 & \textbf{92.10} & \textbf{95.56} & 88.22 & 93.86 & 83.43 & 86.03 & \textbf{85.53} & \textbf{90.05} & \textbf{87.38 $\pm$ 2.90} & 90.59 $\pm$ 3.65 \\
WORD
& 86.48 & 87.92 & 86.65 & 91.17 & \textbf{89.96} & \textbf{94.95} & 78.88 & 81.50 & 83.04 & 86.52 & 85.00 $\pm$ 3.76 & 88.41 $\pm$ 4.52 \\
TotalSegentator
& \textbf{88.32} & \textbf{89.82} & 87.11 & 91.24 & 84.67 & 90.66 & \textbf{92.93} & \textbf{95.89} & 83.16 & 87.77 & 87.24 $\pm$ 3.37 & \textbf{91.08 $\pm$ 2.68} \\
\midrule
Mean $\pm$ SD
& 87.35 $\pm$ 0.69 & 88.19 $\pm$ 0.96 & 84.76 $\pm$ 7.02 & 88.61 $\pm$ 7.24 & 87.02 $\pm$ 2.17 & 92.46 $\pm$ 1.98 & 81.75 $\pm$ 7.74 & 84.37 $\pm$ 7.91 & 83.99 $\pm$ 1.00 & 87.95 $\pm$ 1.30 & 84.97 $\pm$ 5.21 & 88.32 $\pm$ 5.56 \\
\bottomrule
\end{tabular}
}
\caption{Performance comparison of STU-Net-S trained individually on different datasets and evaluated across various CT datasets. Displayed values represent average DSC (\%) and NSD (\%) for the shared classes across these datasets. The 'Mean \(\pm\) SD' values, summarized both per row and per column, reflect the model's variability across various training and testing conditions.}
\label{tab:stu_net_s_performance}
\end{table*}

\begin{table*}[htbp]
\centering
\resizebox{\textwidth}{!}{
\begin{tabular}{@{} l c c c c c c c c c c c c @{}}
\toprule
\multirow{2}{*}{\centering Train/Test} & \multicolumn{2}{c}{FLARE22} & \multicolumn{2}{c}{AMOS CT} & \multicolumn{2}{c}{WORD} & \multicolumn{2}{c}{TotalSegentator} & \multicolumn{2}{c}{BTCV} & \multicolumn{2}{c}{Mean $\pm$ SD} \\
\cmidrule(lr){2-3} \cmidrule(lr){4-5} \cmidrule(lr){6-7} \cmidrule(lr){8-9} \cmidrule(lr){10-11} \cmidrule(lr){12-13}
& DSC & NSD & DSC & NSD & DSC & NSD & DSC & NSD & DSC & NSD & DSC & NSD \\
\midrule
FLARE22
& 87.52 & 87.76 & 72.11 & 75.47 & 85.70 & 90.63 & 73.14 & 75.28 & 84.94 & 88.05 & 80.68 $\pm$ 6.64 & 83.44 $\pm$ 6.66 \\
AMOS CT
& 87.84 & 88.22 & \textbf{92.57} & \textbf{96.23} & 88.28 & 93.92 & 85.80 & 88.34 & \textbf{85.83} & \textbf{90.65} & 88.06 $\pm$ 2.47 & 91.47 $\pm$ 3.15 \\
WORD
& 86.41 & 87.85 & 86.92 & 91.70 & \textbf{90.41} & \textbf{95.28} & 80.53 & 83.34 & 83.62 & 87.56 & 85.58 $\pm$ 3.32 & 89.15 $\pm$ 4.05 \\
TotalSegentator
& \textbf{88.93} & \textbf{90.23} & 88.79 & 93.04 & 85.23 & 91.29 & \textbf{93.01} & \textbf{95.75} & 85.01 & 89.88 & \textbf{88.19 $\pm$ 2.93} & \textbf{92.04 $\pm$ 2.16} \\
\midrule
Mean $\pm$ SD
& 87.68 $\pm$ 0.90 & 88.52 $\pm$ 1.01 & 85.10 $\pm$ 7.77 & 89.11 $\pm$ 8.05 & 87.41 $\pm$ 2.09 & 92.78 $\pm$ 1.90 & 83.12 $\pm$ 7.27 & 85.68 $\pm$ 7.45 & 84.85 $\pm$ 0.79 & 89.04 $\pm$ 1.27 & 85.63 $\pm$ 5.17 & 89.02 $\pm$ 5.51 \\
\bottomrule
\end{tabular}
}
\caption{Performance comparison of STU-Net-B trained individually on different datasets and evaluated across various CT datasets. Displayed values represent average DSC (\%) and NSD (\%) for the shared classes across these datasets. The 'Mean \(\pm\) SD' values, summarized both per row and per column, reflect the model's variability across various training and testing conditions.}
\label{tab:stu_net_b_performance}
\end{table*}

\begin{table*}[ht]
\centering
\resizebox{\textwidth}{!}{
\begin{tabular}{@{} l c c c c c c c c c c c c @{}}
\toprule
\multirow{2}{*}{\centering Train/Test} & \multicolumn{2}{c}{FLARE22} & \multicolumn{2}{c}{AMOS CT} & \multicolumn{2}{c}{WORD} & \multicolumn{2}{c}{TotalSegentator} & \multicolumn{2}{c}{BTCV} & \multicolumn{2}{c}{Mean $\pm$ SD} \\
\cmidrule(lr){2-3} \cmidrule(lr){4-5} \cmidrule(lr){6-7} \cmidrule(lr){8-9} \cmidrule(lr){10-11} \cmidrule(lr){12-13}
& DSC & NSD & DSC & NSD & DSC & NSD & DSC & NSD & DSC & NSD & DSC & NSD \\
\midrule
FLARE22
& 89.20 & 90.19 & 76.53 & 80.25 & 85.94 & 90.76 & 74.06 & 76.56 & 86.11 & 89.28 & 82.37 $\pm$ 5.94 & 85.41 $\pm$ 5.85 \\
AMOS CT
& 89.14 & 89.49 & \textbf{93.02} & \textbf{96.47} & 89.01 & 94.82 & 86.39 & 89.28 & 86.84 & 91.65 & 88.88 $\pm$ 2.35 & 92.34 $\pm$ 2.87 \\
WORD
& 86.86 & 88.73 & 87.53 & 92.34 & \textbf{90.92} & \textbf{95.75} & 80.58 & 83.47 & 84.69 & 88.74 & 86.12 $\pm$ 3.42 & 89.81 $\pm$ 4.10 \\
TotalSegentator
& \textbf{90.32} & \textbf{91.96} & 89.65 & 94.02 & 86.30 & 92.46 & \textbf{95.12} & \textbf{97.33} & \textbf{87.73} & \textbf{92.72} & \textbf{89.82 $\pm$ 3.00} & \textbf{93.70 $\pm$ 1.94} \\
\midrule
Mean $\pm$ SD
& 88.88 $\pm$ 1.26 & 90.09 $\pm$ 1.20 & 86.68 $\pm$ 6.18 & 90.77 $\pm$ 6.25 & 88.04 $\pm$ 2.04 & 93.45 $\pm$ 1.96 & 84.04 $\pm$ 7.74 & 86.66 $\pm$ 7.63 & 86.34 $\pm$ 1.11 & 90.60 $\pm$ 1.64 & 86.80 $\pm$ 4.88 & 90.31 $\pm$ 5.07 \\
\bottomrule
\end{tabular}
}
\caption{Performance comparison of the STU-Net-L model trained individually on different datasets and evaluated across various CT datasets. Displayed values represent average DSC (\%) and NSD (\%) for the eight shared classes across these datasets. The 'Mean $\pm$ SD' values, summarized both per row and per column, reflect the model's variability across various training and testing conditions.}
\label{tab:stu_net_l_performance}
\end{table*}

\begin{table*}[htbp]
\centering
\resizebox{\textwidth}{!}{
\begin{tabular}{@{} l c c c c c c c c c c c c @{}}
\toprule
\multirow{2}{*}{\centering Train/Test} & \multicolumn{2}{c}{FLARE22} & \multicolumn{2}{c}{AMOS CT} & \multicolumn{2}{c}{WORD} & \multicolumn{2}{c}{TotalSegentator} & \multicolumn{2}{c}{BTCV} & \multicolumn{2}{c}{Mean $\pm$ SD} \\
\cmidrule(lr){2-3} \cmidrule(lr){4-5} \cmidrule(lr){6-7} \cmidrule(lr){8-9} \cmidrule(lr){10-11} \cmidrule(lr){12-13}
& DSC & NSD & DSC & NSD & DSC & NSD & DSC & NSD & DSC & NSD & DSC & NSD \\
\midrule
FLARE22
& 88.09 & 89.13 & 71.42 & 75.14 & 86.61 & 91.42 & 72.56 & 74.88 & 85.09 & 88.75 & 80.75 $\pm$ 7.23 & 83.86 $\pm$ 7.29 \\
AMOS CT
& 89.37 & 90.06 & \textbf{92.78} & \textbf{96.46} & 88.85 & 94.61 & 86.13 & 88.82 & 87.00 & 91.76 & 88.83 $\pm$ 2.30 & 92.34 $\pm$ 2.83 \\
WORD
& 85.85 & 87.61 & 87.26 & 92.21 & \textbf{90.93} & \textbf{95.71} & 79.94 & 82.67 & 84.58 & 88.72 & 85.71 $\pm$ 3.58 & 89.38 $\pm$ 4.40 \\
TotalSegentator
& \textbf{90.08} & \textbf{91.75} & 89.58 & 93.91 & 86.45 & 92.62 & \textbf{95.19} & \textbf{97.07} & \textbf{87.44} & \textbf{92.49} & \textbf{89.75 $\pm$ 3.03} & \textbf{93.57 $\pm$ 1.88} \\
\midrule
Mean $\pm$ SD
& 88.35 $\pm$ 1.61 & 89.64 $\pm$ 1.50 & 85.26 $\pm$ 8.23 & 89.43 $\pm$ 8.39 & 88.21 $\pm$ 1.83 & 93.59 $\pm$ 1.67 & 83.46 $\pm$ 8.31 & 85.86 $\pm$ 8.14 & 86.03 $\pm$ 1.22 & 90.43 $\pm$ 1.71 & 86.26 $\pm$ 5.68 & 89.79 $\pm$ 5.92 \\
\bottomrule
\end{tabular}
}
\caption{Performance comparison of STU-Net-H trained individually on different datasets and evaluated across various CT datasets. Displayed values represent average DSC (\%) and NSD (\%) for the shared classes across these datasets. The 'Mean \(\pm\) SD' values, summarized both per row and per column, reflect the model's variability across various training and testing conditions.}
\label{tab:stu_net_h_performance}
\end{table*}

\section{Comprehensive Results of A-Eval}
We present a thorough and comprehensive performance summary of our STU-Net-L model on A-Eval in Table \ref{tab:comprehensive_model_performance}. The model was trained under a variety of conditions and data usage scenarios, including Pseudo Labeling, multi-modality training, and joint training, and subsequently evaluated across diverse datasets, encompassing the MR modality. These extensive training and validation methodologies form a fundamental part of our A-Eval benchmark.

As shown in Table \ref{tab:comprehensive_model_performance}, it can be observed that models trained exclusively on CT datasets, including AMOS CT, perform poorly when tested on the AMOS MR dataset. Conversely, the same phenomenon is seen when models trained solely on the AMOS MR dataset are evaluated using CT datasets. Additionally, models yield the highest performance when trained and evaluated on the same dataset. The only comparable performance is observed from models subjected to joint training. However, there is a noticeable drop in performance when models are trained and evaluated on different datasets. These results not only underscore the challenges of cross-dataset generalizability in abdominal multi-organ segmentation but also indicate the potential viability and effectiveness of joint training strategies in addressing this issue.

\section{Detail Results for Different Model Sizes}
We evaluated the cross-dataset generalizability of four different sized variants of the STU-Net model. The detail results for each model are presented in their corresponding tables: STU-Net-S results in Table \ref{tab:stu_net_s_performance}, STU-Net-B results in Table \ref{tab:stu_net_b_performance}, STU-Net-L results in Table \ref{tab:stu_net_l_performance}, and finally, STU-Net-H results in Table \ref{tab:stu_net_h_performance}. Comparing the results in these four tables, it is apparent that different models under the same training conditions may experience an increase in performance on certain validation sets, but a decrease on others. This underscores the instability of traditional singular validation results. Our A-Eval approach employs cross-dataset validation, utilizing multiple datasets, which can provide more generalized results to some extent.

However, examining the overall trends, we see that model performance increases steadily from STU-Net-S to STU-Net-B to STU-Net-L. However, when we move to STU-Net-H, the performance results are mixed, with some scores showing further improvement and others declining. This suggests that the STU-Net-H model, with its large parameter count of 1.4B, exhibits higher data requirements. Training such a large model from scratch on smaller datasets could potentially lead to suboptimal performance, even falling short of the performance achieved by the more manageable STU-Net-L model. 

\end{document}